\documentclass[prl,12pt,floatfix]{revtex4}
\usepackage{graphicx}
\usepackage{amsmath}

\begin{document}

\title{Cluster growing process and a sequence of magic numbers.}

\author{Andrey Koshelev, Andrey Shutovich }
\affiliation{A.F.Ioffe Physical-Technical Institute,
Politechnicheskaya 26, 194021 St.Petersburg, Russia}

\author{Ilia A. Solov'yov}
\altaffiliation[Permanent address:]{A. F. Ioffe Physical-Technical
Institute of the Russian Academy of Sciences, Polytechnicheskaya 26,
St. Petersburg, Russia 194021}
\email[Email address: ]{solovyov@th.physik.uni-frankfurt.de}
\author{Andrey V. Solov'yov}
\altaffiliation[Permanent address:]{A. F. Ioffe Physical-Technical
Institute of the Russian Academy of Sciences, Polytechnicheskaya 26,
St. Petersburg, Russia 194021}
\email[Email address: ]{solovyov@th.physik.uni-frankfurt.de}
\author{Walter Greiner}
\affiliation{Institut f\"{u}r Theoretische Physik der Universit\"{a}t
Frankfurt am Main, Robert-Mayer 8-10, Frankfurt am Main, Germany 60054}

\begin{abstract}
We present a new theoretical framework for 
modelling the cluster growing process.
Starting from the initial tetrahedral cluster configuration,
adding new atoms to the system and absorbing its energy  
at each step, we find cluster growing paths up
to the cluster sizes of more than 100 atoms. We demonstrate that
in this way all known global minimum structures
of the Lennard-Jonnes (LJ) clusters can be found.
Our method provides an efficient tool for the calculation
and analysis of atomic cluster structure. With its use
we justify the magic numbers sequence for the clusters
of noble gases atoms and compare it with experimental
observations. 
We report the striking correspondence of the peaks in
the dependence on cluster size of the second derivative of
the binding energy per atom calculated for the
chain of LJ-clusters based on the icosahedral symmetry
with the peaks in the abundance mass spectra experimentally measured
for the clusters of noble gases atoms.
Our method serves an efficient alternative to the global optimization techniques
based on the Monte-Carlo simulations and it can be applied for the solution
of a broad variety of problems in which atomic cluster structure  is
important.

\end{abstract}


\maketitle

It is well known that the sequence
of cluster magic numbers carries essential
information about the electronic
and ionic structure of the cluster \cite{LesHouches}. 
Understanding of the 
the cluster magic numbers is often equivalent
or nearly equivalent to the understanding of
cluster electronic and ionic structure. A good example
of this kind is the observation of the magic numbers
in the  mass spectrum of sodium clusters
\cite{Knight84}. In this case, the magic numbers  
were explained by the delocalised electron shell closings 
(see \cite{MetCl99} and references therein).
Another example is the 
the discovery of fullerenes, and in particular the $C_{60}$ molecule
\cite{Kroto85}, 
which was made by means of the carbon clusters mass spectroscopy.

The formation of a sequence of cluster magic
numbers should be closely connected to the mechanisms
of cluster formation and  growing.
It is natural to expect that one can explain
the magic numbers sequence and find the most stable cluster isomers
by modelling mechanisms of  cluster assembling and growing.
On the other hand, these mechanisms are of interest on their own and
the correct sequence of the magic numbers found in such a simulation
can be considered as a proof of validity of 
the cluster formation model.

The problem of magic clusters  is closely connected to the
problem of searching for global minima on the cluster multidimentional
potential energy surface. The number of local minima
on the potential energy surface increases exponentially with the
growth cluster size  and is estimated
to be of the order of $10^{43}$ for $N=100$ \cite{LesHouches}. 
Thus, searching for global minima
becomes increasingly difficult problem for large clusters.
There are different algorithms and methods of the
global minimisation, which have been employed for the global 
minimisation of atomic cluster systems (see \cite{LesHouches} and references
therein). These techniques are often based on the Monte-Carlo simulations.

The algorithm which we describe in this work is based on 
the dynamic searching
for the most stable cluster isomers in the cluster growing process.
Our calculations demonstrate that our approach is 
an efficient alternative to the known techniques of the cluster 
global minimisation. The big advantage of our approach consists in 
the fact that it allows to study  not just the optimized cluster geometries, 
but also their  formation mechanisms.

In the present work we approach the formulated problem 
in a most simple, but general form.  In our most
simple scenario, we assume that atoms in a cluster are bound 
by Lennard-Jones potentials
and the cluster growing  takes place atom by atom.
In this process, new atoms are placed on the cluster surface in the
middle of the cluster faces. Then, all atoms in the system
are allowed to move, while the energy of the system is
decreased. The motion of the atoms is stopped, when the
energy minimum  is reached. The geometries and energies 
of all cluster isomers found in this way are stored
and analysed. The most stable cluster configuration (cluster isomer) 
is then used  as a starting  configuration
for the next step of the cluster growing process.

Starting from the initial tetrahedral cluster configuration
and using the strategy described above, we have analysed 
cluster growing paths up
to the cluster sizes of more than $100$ atoms. We have found that
in this way practically all known global minimum structures
of the Lennard-Jonnes clusters (see \cite{LesHouches} and references
therein) can be determined, which
proves that our method is indeed an efficient 
alternative to other cluster global optimization techniques such
as basin hoping algorithm \cite{LesHouches}. 

In our model we consider an atomic cluster as a group of atoms 
that interact with  each other by pairing forces.  
The interaction potential 
between two atoms in the cluster can, in principle,  be arbitrary.
In this work, we use the  Lennard-Jones (LJ) potential: 
\begin{equation}
U(r) = 4\varepsilon\{ (\sigma/r)^{12} - (\sigma/r)^6\},
\label{LJ_potential}
\end{equation}
where $r$ is the interatomic distance, $-\varepsilon$ is the
depth of the potential well ($\varepsilon > 0$), 
$2^{1/6} \sigma$ is the pair bonding length.


The constants in the potentials allow one to model
various types of clusters for which LJ paring force approximation
is reasonable. The most natural systems of this kind are
the clusters consisting of noble gases atoms Ne, Ar, Kr, Xe.
The magic numbers for this type of clusters 
have been experimentally determined in \cite{Echt81,Haberland94}. 
In our
modelling of the cluster growing process we  focus on this 
example and consider it below in detail.
The constants in the LJ potential appropriate for the noble gases
atoms one can find in \cite{RS}. 
The LJ forces are also appropriate for modelling nuclear clusters
consisting of alpha particles \cite{BM}.
Note that for the LJ clusters it is always possible to chose 
the coordinate scale so that $\sigma = 1$. 
It makes all LJ cluster systems scalable.
They differ only by the choice of the energy parameter $\varepsilon$ and
the mass of a single constituent (atom).

In our approach the atomic motion in the cluster  is described
by the Newton equations with the LJ pairing forces. The system
of coupled equations for all atoms in the cluster are solved numerically
using the 4-th order Runge-Kutta method.
The primary goal in this simulation was to find the solutions
of the equations that lead to the stable cluster configurations
and then to chose energetically the most favourable one.
The choice of initial conditions for the simulation and the algorithm 
for the solution of this problem are described below.

Our cluster searching algorithm is constructed on the idea that 
each minimum on the cluster potential energy surface corresponds to
the situation, when all the atoms are located in their equilibrium
positions. A minimum can be found
by allowing atoms to move, starting from a certain initial  cluster
configuration, and by absorbing all  their
kinetic energy in the most efficient way. 
If the starting cluster configuration for $N+1$ atoms
has been chosen on the basis of the global minimum
structure for $N$ atoms, then it is natural to assume, and 
we prove this in the present work, that the global minimum
structure for $N+1$ atoms can be easily found. The success
of this procedure reflects the fact that in nature 
clusters in their global minima often emerge namely in the
cluster growing process, which we simulate in such calculation.

We have employed the following algorithm for the kinetic energy
absorption. At each step of the calculation
we consider  the motion 
of one atom only, which undergoes the action of the maximum
force.  At the point, in which the kinetic energy
of the selected atom is maximum, we set the absolute
value of its velocity to zero. This point corresponds
to the minimum of the potential well at which the selected atom moves.
When the selected atom is brought to the equilibrium position, 
the next atom is selected to move and  the
procedure of the kinetic energy absorption repeats.
The calculation stops when all the atoms are in equilibrium.

We have considered a number of scenario of the cluster
growing  on the basis of the developed algorithm 
for finding the stable cluster configurations.  

In the most simple scenario  clusters of $N+1$ atoms are generated
from the N-atomic clusters by adding one atom to the system.
In this case the initial conditions for the
simulation of  N+1-atomic clusters 
are obtained on the basis of the chosen N-atomic cluster configuration by
calculating the coordinates of an extra atom added to the system
on a certain rule. We have probed the following paths: 
the new atom can be added either
({\it A1}) to the center of mass of the cluster, or 
({\it A2}) randomly outside the cluster, 
but near its surface, or
({\it A3}) to the centers of mass of 
all the faces of the cluster 
(here, the cluster is considered 
as a polyhedron), 
or 
({\it A4}) to the points that are close 
to the centers of all the faces of the cluster, located 
from both sides of the face on the perpendicular to it,
({\it A5}) to the centers of mass of the faces laying on the cluster
surface. 

The choice of the method how to add atoms to the system depends
on the problem to be solved. For example, 
with {\it A1}  and {\it A2} methods large clusters consisting of 
many particles can be generated rather quickly. The {\it A2} method 
is especially fast, because adding  one atom to the boundary 
of the cluster usually does not lead to the recalculation of its central 
part. The {\it A3} and {\it A4} methods can be used
for searching the most stable, i.e. energetically favourable, 
cluster configurations or for finding  cluster isomers with some
other specific properties.
The {\it A4} method leads to finding more cluster isomers than 
the {\it A3} one, 
but it takes more CPU time. 
The {\it A5} method is especially convenient for modelling the
cluster growing process  which we focus on in this paper. Using this
method one can generate the cluster growing paths for the
most stable cluster isomers.

When considering the cluster growing process, new atoms should be
added to the system starting 
from the initially chosen cluster configuration
step by step until the desired
cluster size is reached. 
Each new step of the
cluster growing should be made with the use
of the methods {\it A1-A5}.
The criteria for the cluster selection in this process can be
as follows: at every step
({\it SE1})  one of the clusters with 
the minimum number of atoms is selected, or
({\it SE2}) the cluster with the minimum energy 
among the already found stable clusters of the maximum size is selected, or  
({\it SE3}) the cluster with the maximum energy 
among the already found stable clusters of the maximum size is selected.  

The {\it SE1} criterion 
is relevant in the situation, when the full search of
cluster isomers is needed. It is
applicable to the systems with relatively
small number of particles.
The {\it SE2} criterion is relevant for modelling
the cluster growing process. It turns out to be very efficient and leads
to finding the most stable cluster configurations
for a given number of particles. The {\it SE3} criterion
might be useful for the redirection of the
cluster growing process towards the lower energy
cluster isomers branches.


Calculations performed with the use of the methods
described above show that often clusters of higher
symmetry group possess relatively low energy. Thus, the symmetric
cluster configurations are often of particular interest.
The process of searching the symmetric cluster configurations
can be speed up significantly, if one performs the cluster
growing process with the imposed symmetry constraints.
This means that for obtaining a symmetric 
$N$ atomic cluster isomer
from the initially chosen symmetric $(N-M)$-atomic configuration one
should add $M$ atoms to the surface of this isomer symmetrically.

\begin{figure}
\begin{center}
\includegraphics[scale=0.57]{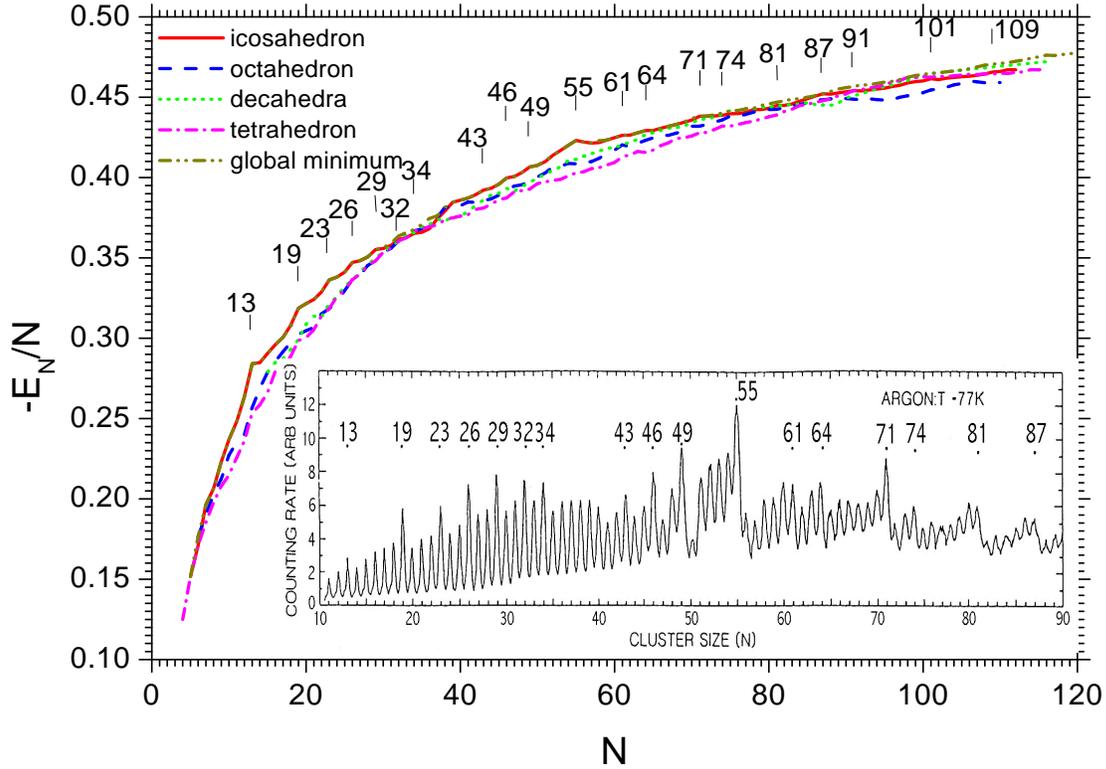}
\end{center}
\caption{
Binding energy per atom for LJ-clusters as a function of
cluster size calculated for the cluster chains 
based on the icosahedral, octahedral, tetrahedral and decahedral 
symmetry. In the insertion we present the experimentally measured abundance 
mass spectrum for the Ar clusters at 77K \cite{Echt81,Haberland94}
}
\label{energies_cl}
\end{figure}

Using our algorithms  we have examined various paths 
of the cluster growing process 
and determined the most stable isomers up to the cluster
sizes of more than 100 atoms.
The binding energies per atom as a function of cluster size  for
the calculated cluster chains are shown in figure
\ref{energies_cl}. In the insertion to figure \ref{energies_cl} 
we present the experimentally measured abundance 
mass spectrum for the Ar clusters at 77K \cite{Echt81,Haberland94}.

We have generated the chains of clusters 
based on the icosahedral, octahedral, tetrahedral and decahedral
symmetries with the use of the  {\it A3-A5} and {\it SE2-SE3}
methods. In a few particular cases for $N > 70$, we have also used
manual modifications of the starting cluster geometries.
In all our calculations we have used the dimensionless 
form of the LJ-potential,
i.e. put $\sigma =1$.
The potential constant has been chosen as 
$\varepsilon = 1/4$.
Such a choice of constants is the most universal. 
It allows one to rescale easily all the results to any concrete choice
of $\sigma$ and $\varepsilon$.

Figure \ref{energies_cl} shows that the most stable clusters 
are obtained on the basis of the icosahedral symmetry configurations
with exceptions for $N=38$, $75 \leq N \leq 77$ and $N=98$. In
these cases the octahedral cluster symmetry becomes more favourable. 
The cluster chains based on the tetrahedral and decahedral symmetries
have no intersections with the icosahedral chain of clusters although
there is interplay between these two curves and the octahedral one.

\begin{figure}
\begin{center}
\includegraphics[scale=0.57]{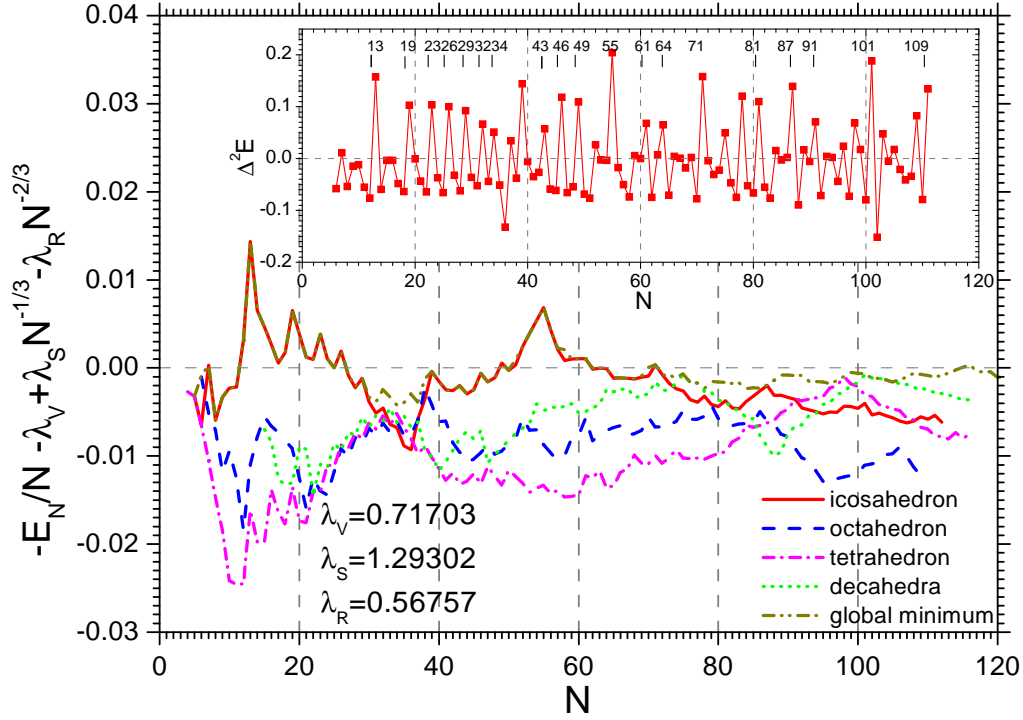}
\end{center}
\caption{Energy curves deviations from the liquid drop
model (\ref{drop_model}) calculated for
various cluster isomers chains.
In the insertion we plot the second derivative 
$\Delta^2 E_N=E_{N+1}+E_{N-1} -2 E_N$ 
calculated for the icosahedral cluster isomers chain. 
}
\label{drop_model}
\end{figure}

The main trend of the energy curves 
plotted in figure \ref{energies_cl} can be
understood on the basis of the liquid drop model,
according to which the cluster energy 
is the sum of the volume and the surface energy contributions:
\begin{equation}
E_{N}= - \lambda_V N + \lambda_S N^{2/3} - \lambda_R N^{1/3}
\label{LD_model}
\end{equation}
Here the first and the second terms describe 
the volume, and the surface
cluster energy correspondingly. The third term is the
cluster energy arising due to the curvature of the cluster surface.
Choosing constants 
in (\ref{LD_model}) as $\lambda_V=0.71703$,  $\lambda_S=1.29302$
and $\lambda_R=0.56757$, one can
fit the global energy minimum curve plotted in figure \ref{energies_cl}
with the accuracy less than one per cent.  The deviations of
the energy curves calculated for
various chains of cluster isomers from the liquid drop
model (\ref{drop_model})
are plotted in figure \ref{drop_model}.
The curves for the icosahedral and the global energy
minimum cluster chains go very close with each other and 
the peaks on these dependences indicate
the increased stability of the corresponding magic clusters.
The ratio between the volume and surface energies
in (\ref{LD_model}) can be characterised by the dimensionless parameter
$\delta=\lambda_V/\lambda_S$, being equal in
our case to $\delta=0.5545$. 

The dependence of the binding energies per atom for the most stable cluster
configurations on $N$ allows one to generate the sequence of the
cluster magic numbers. In the insertion to
figure \ref{drop_model} we plot the second
derivatives $\Delta E_n^2$
for the chain of icosahedral
isomers.
We compare the obtained dependence with the experimentally measured abundance 
mass spectrum
for the Ar clusters at 77K \cite{Echt81,Haberland94} (see insertion to
figure \ref{energies_cl}) and
establish the striking correspondence
of the peaks in the measured mass spectrum with those
in the $\Delta E_n^2$ dependence. 
Indeed, the magic numbers determined from $\Delta^2 E_N$
are in a very good
agreement with the numbers experimentally measured 
for the Ar and Xe clusters: 
13, 19, 23, 26, 29, 32, 34, 43, 46, 49,
55, 61, 64, 71, 74, 81, 87,  91,  101, 109, 116, 119, 124, 131, 136, 147
\cite{Echt81,Haberland94}.
The most prominent peaks in this sequence 13, 55 and 147 
correspond to the closed icosahedral
shells, while other numbers correspond to the filling of
various parts of the isosahedral shell. 

The connection between the second derivatives $\Delta^2 E_N$
and the peaks in the abundance mass spectrum of clusters
one can understand using the following simple model.
Let us assume that the mass spectrum of clusters is formed
in the evaporation process. This means that changing
the number of clusters, $n_N$, of the size $N$ in the
cluster ensemble takes place due to the evaporation of an atom by the clusters
of the size $N$ and $N+1$, i.e.
$\Delta n_N \sim n_{N+1} W_{N+1 \rightarrow N} - n_{N} W_{N \rightarrow N-1}$,
where the evaporation probabilities  are proportional
to $W_{N+1 \rightarrow N} \sim e^{-\frac{E_N+E_1-E_{N+1}}{kT}}$
and $W_{N \rightarrow N-1} \sim e^{-\frac{E_{N-1}+E_1-E_N}{kT}}$. Here
$T$ is the cluster temperature,
$k$ is the Bolzmann constant. In the limit $\Delta E_N/kT \ll 1$, one
derives $ \Delta n_N \sim n_N (E_{N+1}+E_{N-1} -2 E_N)/kT \sim \Delta^2 E_N$.
These estimates demonstrate that the positive second derivative 
$\Delta^2 E_N$ should lead to the enhanced abundance of 
the corresponding clusters.

\begin{figure}
\begin{center}
\includegraphics[scale=0.72]{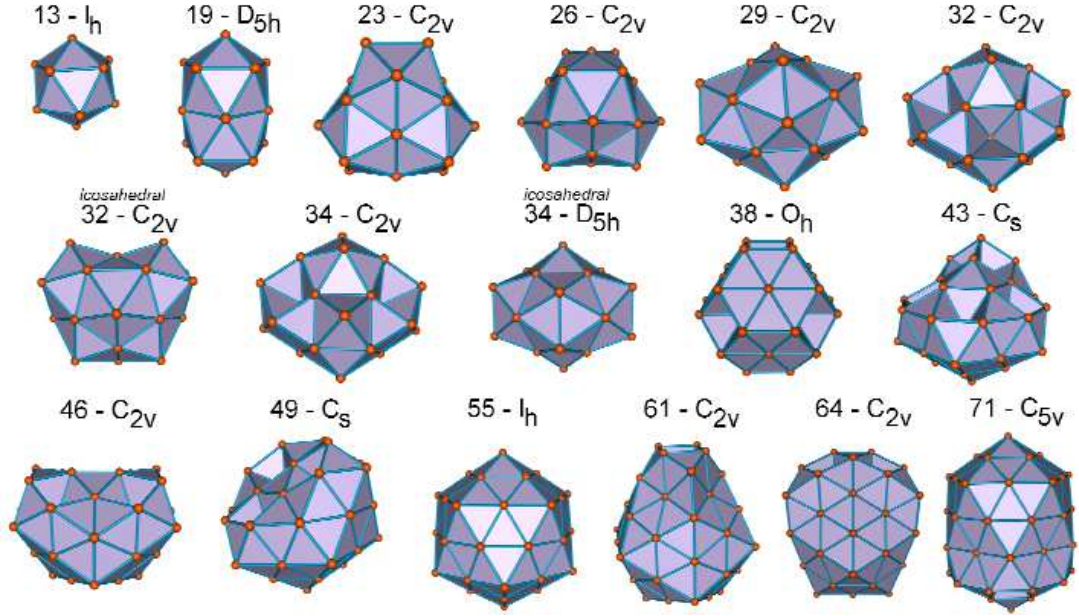}
\end{center}
\caption{
Geometries of the magic LJ-clusters. 
The labels indicate the cluster size and
the cluster point symmetry group.
}
\label{cl_images}
\end{figure}

In figure \ref{cl_images}, we plot  images of the magic
clusters up to $N=71$. 
For $N=32$ and $N=34$, we present the icosahedral isomer
and the one possessing the global energy minimum.
We also plot the image of
the octahedral $N=38$ cluster, which is found to be more stable
than the clusters from the icosahedral chain.
Experimentally $N=38$ is not found to be 
the magic cluster, although it is
the global minimum cluster, being magic
for the octahedral cluster chain  (see 
figures \ref{energies_cl} and \ref{LD_model}). 
This fact can be understood if one takes into account
that different symmetry cluster chains are formed independently
and the transition of clusters
from one chain to another at certain $N$ is not possible.
It is clear from the
binding energy analysis that the icosahedral chain of clusters
should be dominating. In experiments, clusters of
the icosahedral chain mask clusters of other symmetry
chains even in the situations when these other clusters
are energetically more favourable, like it occurs for $N=38$.

In this paper we have discussed the classical models for
the cluster growing process, but  our ideas 
can be easily generalized on the quantum case and
be applied to the cluster systems with different than LJ type 
of the inter-atomic interaction. It would be interesting
to see to which extent the parameters of  inter-atomic interaction
can influence the cluster growing process and the corresponding
sequence of  magic numbers or whether the crystallization
in the nuclear matter consisting of alpha particles and/or nucleons 
is possible. Studying cluster thermodynamic
characteristics with the use of the developed
technique is another interesting problem which is left opened for
future considerations.


The authors acknowledge support from the
the Alexander von Humboldt Foundation and
DAAD.

\end{document}